# A Hybrid Prior Bayesian Method for Combining Domestic Real-World Data and Overseas Data in Global Drug Development

**Keer Chen [1], Zengyue Zheng [1], Pengfei Zhu [2], Shuping Jiang [2], Nan Li [3], Jumin Deng [1], Pingyan Chen [4], Zhenyu Wu [5*], Ying Wu [1,4*]**

[1] Department of Biostatistics, School of Public Health, Southern Medical University, Guangzhou 510515, China

[2] BARDS, MSD China, Shanghai, China

[3] BARDS, MSD R&D (China) Co., Ltd., Beijing, China

[4] Hainan Institute of Real World Data, The Administration of Boao Lecheng International Medical Tourism Pilot Zone, Hainan 571434, China

[5] Department of Biostatistics, School of Public Health, Fudan University, No.130 Dong-An Road, Shanghai, 200032, China. Tel/fax: +86 (21) 54237707. E-mail: zyw@fudan.edu.cn;

[*]Corresponding to: Ying Wu (wuying19890321@gmail.com) and Zhenyu Wu

# Abstract

## Background

Hybrid clinical trial design integrates traditional randomized controlled trials (RCTs) with real-world data (RWD), aiming to enhance trial efficiency through dynamic incorporation of external data (External trial data and RWD). However, existing methods, such as the Meta-Analytic Predictive Prior (MAP), exhibit significant limitations in controlling data heterogeneity, adjusting baseline discrepancies, and optimizing dynamic borrowing proportions. These limitations often introduce external bias or compromise evidence reliability, hindering their application in complex analyses like bridging trials and multi-regional clinical trials (MRCTs).

## Objective

This study proposes a novel hybrid Bayesian framework, EQPS-rMAP, to address heterogeneity and bias in multi-source data integration. Its feasibility and robustness are validated through systematic simulations and retrospective case analyses, using two independent datasets evaluating risankizumab's efficacy in patients with moderate-to-severe plaque psoriasis.

**Design and Methods**

The EQPS-rMAP method operates in three stages: (1) eliminating baseline covariate discrepancies through propensity score stratification; (2) constructing stratum-specific MAP priors to dynamically adjust weights for external data; and (3) introducing equivalence probability weights to quantify data conflict risks. The study evaluates the method's performance across six simulated analyses (heterogeneity differences, baseline shifts, etc.), comparing it with traditional methods (MAP, PSMAP, EBMAP) in terms of estimation bias, type I error control, and sample size requirements. Real-world case analyses further validate its applicability.

**Results**

Simulations demonstrate that EQPS-rMAP maintains estimation robustness under significant heterogeneity while reducing sample size demands and enhancing trial efficiency. Case analyses confirm its ability to control external bias while preserving high estimation accuracy compared to conventional approaches.

**Conclusion and Significance**

The EQPS-rMAP method provides empirical evidence for the feasibility of hybrid clinical designs. Its methodological advancements—resolving baseline and heterogeneity conflicts through adaptive mechanisms—offer broader applicability for integrating external and real-world data across diverse analyses, including bridging trials, MRCTs, and post-marketing studies.

## 1 Introduction

In the domain of confirmatory research for novel drug development, randomized controlled trials (RCTs) remain the gold standard for efficacy evaluation. However, strategic integration of external data resources has emerged as a critical pathway to overcome efficiency bottlenecks in R&D. Data from hospital information systems, natural population cohorts, and concurrent/historical clinical trials—containing valuable drug effectiveness and safety information—provide complementary evidence for bridging trials and multi-regional clinical trials (MRCTs). A core paradox persists: while the FDA actively promotes real-world data (RWD) utilization through initiatives like the

Complex Innovative Trial Design Pilot Program[1] and Clinical Trial Transformation Initiative[2], challenges including data heterogeneity, inherent biases, and therapeutic effect variability continue to undermine the reliability of external data integration.

The ICH E5 guideline[3]-established bridging trial framework and widely adopted MRCT paradigm provide regulatory foundations for cross-regional data harmonization. Statistical methodologies for these analyses have evolved along two pathways[4]: classical approaches employing reconstruction probability, weighted Z-tests, and sequential designs[5-12] for data alignment, and Bayesian frameworks incorporating mixed priors[13] and power priors[14-17]. Traditional Bayesian mixed priors address inter-regional information conflicts but suffer from preset weighting constraints that limit practical utility. To overcome this limitation, Meta-Analytic Predictive (MAP) prior methods[18] and derivatives—including Robust MAP[19], Bayesian semiparametric MAP prior[20], empirical Bayesian MAP priors[21], and the Empirical Bayes Robust MAP Prior (EB-rMAP)[22]—dynamically adjust data borrowing proportions based on the consistency between historical and current datasets. However, within frameworks integrating Real-world data (RWD) and External trail data, baseline discrepancies and varying effect heterogeneity across data sources persist. Traditional methods fail to address baseline differences and cannot simultaneously account for heterogeneity variations among multiple data sources. Instead, they simplistically assume homogeneity across sources, applying uniform borrowing proportions.

China's established Real-World Study (RWS) guidelines now provide regulatory frameworks for innovative drug development pathways. The Boao Lecheng Pilot Zone's RWD-supported drug registration exemplifies regulatory innovation. This study introduces EQPS-rMAP, an advanced data-sharing strategy applicable to bridging studies, MRCTs, and post-marketing confirmatory research. The method constructs unified hybrid priors by synthesizing domestic RWD with foreign RCT data, stratifying patient subgroups through baseline covariate matching to create homogenized layers. Within each stratum, layer-specific MAP priors are integrated with optimized weighting to reconcile effect heterogeneity and baseline discrepancies, augmented by uninformative prior weighting for enhanced robustness. Consequently, EQPS-rMAP enables adaptive data borrowing through adjustable threshold parameters, achieving precision-controlled external evidence integration. The following sections describe the theoretical foundations and practical applications of the PS and MAP methods, leading to the introduction of the EQPS-rMAP methodology in Section 2. Section 3 demonstrates the robustness of the EQPS-rMAP through extensive simulation studies across various analyses, and Section 4 discusses the empirical results.

## 2 Statistical Methods

In anticipation of conducting a bridging study within a new geographical region, the data amalgamation encompasses overseas randomized controlled trial data alongside domestic real-world data, which solely encompasses the test group. It is posited that the responder count across all data segments adheres to a one-parameter exponential family of distributions $Y \sim Bin(n,p)$, providing a standardized framework for analysis.

### 2.1 Review of propensity score for incorporating external data

A propensity score encapsulates the likelihood of an individual receiving a particular treatment, accounting the cumulative influence of confounding variables (X) on the treatment decision[23]. This metric is crucial for addressing non-exchangeability issues in causal inference studies. In this research, the propensity score is defined as the probability of an individual's inclusion in Current trial data[24]. Index vector $\boldsymbol{Z} = (Z_1, Z_2)$ is the data source indicator vector with the following assignment rule.

$$\boldsymbol{Z} = \begin{cases} (1,0), & \text{Real} - \text{world data} \\ (0,1), & \text{External trial data} \\ (0,0), & \text{Current trial data} \end{cases}$$

In this paper, we define the propensity score $e(X)$ as the probability that an individual belongs to Current trial data. Propensity score models were constructed based on External trial data, Real-world data and Current trial data. Let the vector $X$ denote all covariates:

$$e(X) = \Pr(Z = (0,0)|X) \#(2-1)$$

The analysis examines the likelihood of a participant coming from a new study area based on primary confounding factors. It emphasizes that participants with similar propensity scores exhibit comparable distributions of baseline covariates, ensuring balanced group comparisons. Logistic regression remains the standard method for calculating propensity scores. However, recent progress advocates alternative estimation techniques such as random forests[25] neural networks[26], and ensemble methods such as bagging or boosting[25, 27].

Propensity score analysis employs strategies like matching, weighting, and stratification for bias reduction. Specifically, stratification divides participants into quantiles based on propensity scores, ensuring homogeneity within each stratum regarding baseline characteristics[27-28]. This process facilitates balanced outcome estimation across strata.

*2.2 Review of Meta-Analytical-Predictive Prior*

Meta-analytic predictive (MAP) prior approaches leverage robust hierarchical modeling frameworks to accommodate varying levels of between-trial heterogeneity. The MAP prior is constructed by deriving the predictive posterior distribution of the target parameter (e.g., treatment effect $\theta$) from external studies and is subsequently integrated with current trial data during the analysis phase of the new study[19].

For instance, consider a single-arm trial designed to estimate parameter $\theta$, where external evidence will be systematically incorporated to enhance the precision of inferences. In general, $d_O = \{Y_1, Y_2, \cdots Y_H\}$ denotes the number of responders in External trial data and $d_* = \{Y_*\}$ denotes the number of responders in Current trial data. $n$ represents the number of patients and $p$ represents the patient response rate. The MAP approach adopts the following hierarchical model[19]. First, suppose the binary response is modeled by

$$i = 1, \cdots H, * \quad Y_i \sim Bin(n_i, p_i)$$

The similarity of Current trial data and External trial data is expressed by exchangeable parameters $\theta$.

$$\theta_*, \theta_1, \cdots, \theta_H \sim Normal(\mu, \tau^2), \text{where } \theta_i = log(\frac{p_i}{1-p_i})$$

For the overall mean $\mu$, a vague prior was used because the data had sufficient information[19]. The intertrial standard variance $\tau$ can be distributed following a half-T distribution[29-30], and in this paper, we utilize a half-normal distribution, with standard deviations selected in such a way that values of $\tau$ that are implausibly large have a low probability of occurring.

Given Current trial data and External trial data, the posterior distribution $g_{MAP}(\theta_*|d_0, d_*)$ of interest can be expressed.

$$g_{MAP}(\theta_*|d_0, d_*) \propto p(d_*|\theta_*) \cdot f_{MAP}(\theta_*|d_0) \#(2-2)$$

Where $p(d_*|\theta_*)$ represents the likelihood function of the Current trial data, $f_{MAP}(\theta_*|d_0)$ is the original MAP prior[18]. In practical applications, the sample size from the original geographical area often exceeds that of the new geographical region. This discrepancy presents a challenge in evaluating the posterior distribution of treatment effects, as the smaller sample size from the new region may not substantially alter the combined outcome. In order to improve the robustness, Robust MAP (rMAP)[19] prior is proposed, which adds a vague prior component $f_v(\theta_*)$ to $f_{MAP}(\theta_*|d_0)$. For example, for binomial distribution, the standard uniform prior Beta(1,1) or Jefferys' prior $Beta(0.5,0.5)$ is generally used.

$$(1 - \omega_v) \cdot f_{MAP}(\theta_*) + \omega_v \cdot f_v(\theta_*) \#(2-3)$$

$\omega_v \epsilon (0,1)$ is used to control the proportion of borrowing from reference data (including External trial and Real-world data) in construction of the rMAP, when $\omega_v = 1$ represents no borrowing. When direct sampling of the MCMC from the unknown distribution is challenging, the authors of this study utilized the mixture approximation of the conjugate prior, as derived by Dalal and Hal[17]. For the binomial distribution, $f_{MAP}(\theta_*)$ can be approximated as a weighted mixture of beta distributions. The Kullback-Leibler scatter is used to measure the components. For example, in the binary case, this mixture prior is[19]

$$\hat{f}_{MAP}(\theta_*|d_0) = \sum_{k=1}^{K} \pi_k Beta(a_k, b_k) \#(2-4)$$

$\sum_{k=1}^{K} \pi_k = 1$, $K$ represents the number of components in the approximation. By combining the above formulas, we arrive at the final expression for the prior. According to Schmidli[19], the best approximation of the exact prior is obtained by choosing mixture weights $\pi_k$, and hyperparameters of the conjugate prior. Nevertheless, the RBesT package in R can now be used to approximate this value.

$$\hat{f}_{rMAP}(\theta_*|d_0) = (1 - \omega_v) \cdot \sum_{k=1}^{K} \pi_k Beta(a_k, b_k) + \omega_v \cdot f_v(\theta_*) \#(2-5)$$

One issue is determining how to calculate the value of the vague prior weight $\omega_v$. In bridging experiments, the weights in the Bayes hybrid prior[13] often depend on the researcher's subjective experience to specify. The Robust MAP prior[19] specifies the vague prior weight $\omega_v$ by first providing an initial value and then updating it through marginal probabilities. This approach adds robustness to the previous method. The Empirical Bayes Robust MAP prior[22] specifies the vague prior weight $\omega_v$ by constructing a step function based on Box's prior predicted p-value, making the computation of the weights data dependent, where the adjustable parameters give the model both objectivity and flexibility.

*2.3 Proposed method*

For the current study which compares a test treatment with a control, our primary analysis of interest is a hypothesis test of $H_0: \theta_c = \theta_t$ against $H_1: \theta_c < \theta_t$. The analysis can be considered a trial success if

$$Pr(\theta_c < \theta_t | current\ data\ and\ prior) > 0.95.$$

Assuming a bridging trial including both an experimental drug group and a control group is going to be conducted in a new region. The reference data from which information can be borrowed for this paper includes data from randomized controlled trials overseas (both treatment and control groups) and Real-world data containing only treatment group.

*Step 1: On-trial Propensity score-based stratification*

Propensity scores were calculated from baseline characteristics of External trial data, Real-world data, and Current trial data (defined in Section 2.1).Let $E_1$ denote the set of propensity scores for Current trial data, and $E_2$ denote the set of propensity scores of Real-world data, and $E_3$ denotes the set of propensity scores of External trial data. Figure 1 illustrates the propensity score trimming and stratification process. In order to ensure higher population consistency, the range of values of $E_1$ was used as the boundary, and data for $E_2$ and $E_3$ lying outside this range were trimmed. The culled data are then uniformly stratified by quantiles of $E_1$. The number of stratums is denoted by $S$, which is typically five and quartiles are used for stratification. Let $q = \{\hat{q}_0, \cdots \hat{q}_S\}$ denote the number of boundary-responsive quartiles of $S$ strata, $n_{Ex,S}, n_{Curr,S}, n_{Real,S}$ denote the number of patients in External trial data, the number of patients in Current trial data, and the number of patients in Real-world data in stratum $S$, respectively.

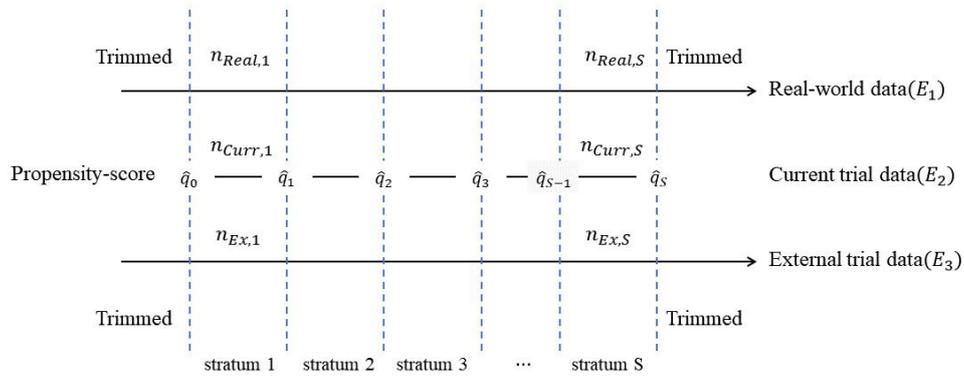

**Figure 1** Illustration of the propensity score trimming and stratification process

*Step 2: Stratum-specific MAP prior*

Perform stratum-specific parameter estimation for each stratum. For simplicity, a binomial distribution is used here to illustrate. In fact, this method can also be easily extended to other outcome distributions such as normal distribution, time-to-event (TTE) endpoint, etc. Assuming the outcome of the experiment $Y$ follows $Y \sim Bin(n, p)$, where $Y$ represents the number of responders in the data, $p$ represents the response rate parameter, and denote $\theta$ to represent $\log(\frac{p}{1-p})$. Let $Y_{Z,s}$ denote the number of responders in stratum $s$ whose trial source is $Z$, where $s = 1, \cdots, S; Z = (1,0), (0,1), (0,0)$ (for example, $Y_{(0,0),1}$ represents the number of responders for the Current trial in the first stratum). $\theta_{Ex,s}$ and $\theta_{Real,s}$ represent parameters within External trial data stratum and Real-world data stratum, respectively. The structure of $Y_{Z,s}$ can be modeled using a Bayesian hierarchical model as follows:

$$Y_{Z,s} \sim Bin(n_{Z,s}, p_{Z,s})$$

$$\theta_{Ex,s} = log(\frac{p_{Z,s}}{1 - p_{Z,s}}) \quad \theta_{Ex,s} | \mu_s, \tau_{Ex,s}^2 \sim Normal(\mu_s, \tau_{Ex,s}^2);$$

$$s = 1, \cdots, S; \mathbf{Z} = (1,0)$$

$$\theta_{Real,s} = log(\frac{p_{Z,s}}{1 - p_{Z,s}}) \quad \theta_{Real,s}|\mu_s, \tau^2_{Real,s} \sim Normal(\mu_s, \tau^2_{Real,s});$$

$$s = 1, \cdots, S; \mathbf{Z} = (0,1) \#(2-11)$$

The stratum-specific MAP prior assumes that the means from different data sources and within different specific stratums originate from the same distribution. A vague prior is used for $\mu_s$ because the data have sufficient information. Heterogeneity between three data source is represented by the variance parameter $\tau^2_{Ex,s}$ and $\tau^2_{Real,s}$, which uses a half-normal prior. Since there are two types of heterogeneity included here, the heterogeneity between stratums and the heterogeneity between different data sources. Our paper adopts $\tau^2_{Ex,s}$ and $\tau^2_{Real,s}$ to denote the heterogeneity between the External trial data and the Real-world data within the same stratum. So in this paper, we recommend selecting the hyperparameters for the prior on $\tau^2_{Ex,s}$ to reflect the magnitude of stratum-specific similarity between each stratum of External trial data and the Current trial data relative to the baseline covariates; and $\tau^2_{Real,s}$ to reflect the magnitude of stratum-specific similarity between each stratum of Real-world data and the Current trial data relative to the baseline covariates. In this paper, we propose to specify a prior in the form of the following hyperparameter, such that the prior can change the proportion of different stratums borrowed in the presence of different similarities in the stratums.

$$\tau_{Ex,s} \sim half-normal(k_{Ex,s}), \quad \tau_{Real,s} \sim half-normal(k_{Real,s})$$

Where $k_{Ex,s}$ and $k_{Real,s}$ reflect the stratum-specific similarities between External trial data and Current trial data, and between Real-world data and Current trial data, respectively, relative to baseline covariates. Then the similarity measure between the two sources of data was expressed according to Wang et al[31] using the overlapping coefficient between the propensity score distributions $f_{Ex,s}(e)$, $f_{Curr,s}(e)$ and $f_{Real,s}(e)$[32].

$$r_{Ex,s} = \int_0^1 \min[f_{Ex,s}(e), f_{Curr,s}(e)] \, de,$$

$$r_{Real,s} = \int_0^1 \min[f_{Real,s}(e), f_{Curr,s}(e)] \, de \#(2-12)$$

The overlap coefficient of stratum $s$, $R_{Ex} = (r_{Ex,1}, r_{Ex,2}, \cdots, r_{Ex,s})$, $R_{Real} = (r_{Real,1}, r_{Real,2}, \cdots, r_{Real,s})$. $k_{Ex,s}$ and $k_{Real,s}$ can then be obtained using the overlap coefficient calculation[33-34].

$$k_{Ex,s} = \frac{r_{Ex}^{ref}}{r_{Ex,s}}, \quad k_{Real,s} = \frac{r_{Real}^{ref}}{r_{Real,s}}$$

Where $r_{Ex}^{ref}$ is generally taken as the median of $R_{Ex}$, and for $r_{Real}^{ref}$ the same is true.

*Step 3: Equivalence-probability-weight based EQPS-rMAP prior*

Let $d_{Ex}$ denotes the External trial data, $d_{Real}$ denotes Real-world data, and $d_{Curr}$ denotes Current trial data, $\theta_{curr}$ denotes the parameter of interest for Current trial data. The stratum-spcific MAP prior was integrated with specific weights into the $f_{MAP}(\theta_{curr}|d_{Real}, d_{Ex})$. Rather than defining an overall EQPS-rMAP prior based on the layer-specific MAP prior, we defined log-odds of the stratum-specific response rate as $\theta_s(s = 1, \dots, S)$ and the log-odds for the overall response rate of Current trial data as $\theta_{curr}$. The EQPS-rMAP prior is then the distribution induced by this relationship. In particular, we define

$$\theta_{curr} = \sum_{S=1} \frac{n_{Curr,s}}{N_{Curr}} * \theta_s$$

$$\theta_s = \omega_{Real,s} * \theta_{Real,s} + \omega_{Ex,s} * \theta_{Ex,s}$$

$$\omega_{Real,s} = \frac{n_{Real,s} * \epsilon_{Real \cdot Curr,s}}{n_{Ex,s} * \epsilon_{Ex \cdot Curr,s} + n_{Real,s} * \epsilon_{Real \cdot Curr,s}},$$

$$\omega_{Ex,s} = \frac{n_{Ex,s} * \epsilon_{Ex \cdot Curr,s}}{n_{Ex,s} * \epsilon_{Ex \cdot Curr,s} + n_{Real,s} * \epsilon_{Real \cdot Curr,s}} \#(2-13)$$

The weights here take into account the consistency of External trial data, Real-world data with Current trial data and the sample size in each stratum. That is, the log-odds for the overall response rate $\theta_{curr}$ of Current trial data is the weighted sum of log-odds of the stratum-specific $\theta_{Real,s}$ of Real-world data and $\theta_{Ex,s}$ of External trial data. Where $n_{Real,s}, n_{Ex,s}, n_{Curr,s}$ denote the sample sizes in stratum $s$ of Real-world data, External trial data, and Current trial data, respectively. $N_{Curr}$ denotes the total sample size of Current trial data. In each stratum, $\epsilon_{Ex\cdot Curr,s}$ indicates the consistency of External trial data with Current trial data, while $\epsilon_{Real\cdot Curr,s}$ indicates the consistency of Real-world data with Current trial data. Consistency is evaluated based on the tail region probability proposed by Thompson[35], calculated as follows,

$$\epsilon_{Ex\cdot Curr,s} = 2 \times min\{Pr(p_{Ex,s} > p_{Curr,s}), 1 - Pr(p_{Ex,s} > p_{Curr,s})\}$$

$$\epsilon_{Real\cdot Curr,s} = 2 \times min\{Pr(p_{Real,s} > p_{Curr,s}), 1 - Pr(p_{Real,s} > p_{Curr,s})\}$$

Where

$$p_{Ex,s} \sim Beta(y_{Ex,s}, n_{Ex,s} - y_{Ex,s}),$$

$$p_{Real,s} \sim Beta(y_{Real,s}, n_{Real,s} - y_{Real,s}),$$

$$p_{Curr,s} \sim Beta(y_{Curr,s}, n_{Curr,s} - y_{Curr,s})$$

Therefore

$$Pr(p_{Ex,s} > p_{Curr,s}) = \int_0^1 \int_{p_{Curr,s}}^1 \frac{p_{Curr,s}^{y_{Curr,s}-1}(1-p_{Curr,s})^{n_{Curr,s}-1}}{B(y_{Curr,s}, n_{Curr,s}-y_{Curr,s})}$$
$$\cdot \frac{p_{Ex,s}^{y_{Ex,s}-1}(1-p_{Ex,s})^{n_{Ex,s}-1}}{B(y_{Ex,s}, n_{Ex,s}-y_{Ex,s})} dp_{Curr,s} dp_{Ex,s} \#(2-14)$$

$y_{Ex,s}$ and $p_{Ex,s}$ denotes the number of responding patients and response rate for each stratum of External trial data, respectively. $y_{Real,s}$ and $p_{Real,s}$ denote the number of responding patients and response rate for each stratum of Real-world data, respectively. $y_{Curr,s}$ and $p_{Curr,s}$ denote the number of responding patients and response rate for each stratum of Current trial data.

Following the established Robust MAP prior framework, this study applies Markov chain Monte Carlo (MCMC) techniques to derive samples from a prior distribution that integrates External trial data with Real-world data. Our approach builds upon the insights of Schmidli et al[19] and Hupf et al[20] and benefits from the use of conjugate priors, which notably enhance density estimation. The number of components, $k$, is up to the number of historical data, and is usually chosen to be the smallest number that provides an adequate approximation based on a particular criterion, e.g. Kullback-Leibler divergence or AIC/BIC. The remaining parameters—mixture weights $\pi_k$ and Beta hyperparameters $a_k$, $b_k$ —are iteratively estimated via the Expectation-Maximization (EM) algorithm, which alternates between computing component assignments (E-step) and maximizing the complete-data likelihood (M-step) [22]. To enhance robustness, we introduce vague prior components that align with the conjugate structure of the model. For example, with binary data, $f_v(\theta_0)$ can be the standard uniform prior $Beta(1,1)$ or Jefferys prior $Beta(0.5,0.5)$. For the prior and posterior of the binomial distribution, it can be expressed in the following form:

$$\hat{f}_{rMAP}(\theta_{curr}|d_{Real}, d_{Ex}) = (1 - \omega_v) \cdot \sum_{k=1}^{K} \pi_k Beta(a_k, b_k) + \omega_v \cdot f_v(\theta_0)$$

$$\tilde{g}_{EQPS}(\theta_{curr}|d_{Real}, d_{Ex}, d_{Curr})$$
$$= (1 - \tilde{\omega}_v) \sum_{k=1}^{K} \pi_k Beta(a_k + y_{Curr}, b_k + N_{Curr} - y_{Curr})$$
$$+ \tilde{\omega}_v Beta(a_0 + y_{Curr}, b_0 + N_{Curr} - y_{Curr}) \#(2-15)\#$$

The weight of the vague prior in this study is determined by the Equivalence Probability Weight indicator[24], assessing the consistency between the mixed posterior distribution, and the current experimental data.

In this study, we define the optimal vague prior weight, denoted as $\omega_{Eq}$, as the smallest prior weight that ensures the consistency metric $p$ meets a predefined threshold $\lambda$. This consistency metric quantifies the agreement between the posterior distribution incorporating external data and the response probability distribution derived from Current trial data.

First, the response probability distribution for the Current trial data is modeled as:
$$p_{Curr} \sim Beta(y_{Curr}, N_{Curr} - y_{Curr})$$

The hybrid posterior distribution, denoted $p_{Hyb}$, is obtained under the EQPS-rMAP framework:
$$p_{Hyb} \sim \tilde{g}_{EQPS}(\theta_{curr}|d_{Real}, d_{Ex}, d_{Curr})$$

The EQPS-rMAP prior is constructed as a mixture of informative components (derived from external data) and a vague prior component. It is expressed as:
$$\hat{f}_{EQPS-rMAP}(\theta_{curr}|d_{Real}, d_{Ex}) = (1 - \omega_{Eq})\sum_{k=1}^{K}\pi_k Beta(a_k, b_k) + \omega_{Eq} \cdot f_v(\theta_{curr}) \quad (2-16)$$

where $f_v(\theta_{curr})$ represents a vague (non-informative) prior, used in cases of high conflict between external and Current trial data.

Upon observing the current trial data, the posterior distribution becomes:
$$\tilde{g}_{EQPS-rMAP}(\theta_{curr}|d_{Real}, d_{Ex}, d_{Curr})$$
$$= (1 - \tilde{\omega}_{Eq})\sum_{k=1}^{K}\pi_k Beta(a_k + y_{Curr}, b_k + N_{Curr} - y_{Curr})$$
$$+ \tilde{\omega}_{Eq} Beta(a_0 + y_{Curr}, b_0 + N_{Curr} - y_{Curr}) \quad (2-17)$$

To assess the agreement between the hybrid posterior and the current data, we define the following probability:

$$p = Pr(p_{Hyb} - \delta < p_{Curr} < p_{Hyb} + \delta) \#(2-18)$$

Here, $\delta$ denotes the clinical equivalence margin, representing the maximum acceptable deviation between the two distributions.

The optimal vague prior weight $\omega_{Eq}$ is then defined as:

$$\omega_{Eq} = \begin{cases} min_{\omega_v}\{\omega_v : p \geq \lambda\}, & if\ \exists \omega_v\ s.t.\ p \geq \lambda \\ 1, & if\ otherwise \end{cases} \#(2-19)$$

This means that if the hybrid posterior is sufficiently consistent with the current data ($p \geq \lambda$), we adopt the smallest possible weight $\omega_{Eq}$ that satisfies this condition, thus maximizing the borrowing of external information. If the consistency is below the threshold ($p < \lambda$), the model defaults to full use of the vague prior (i.e., $\omega_{Eq} = 1$), completely excluding external data.

Both parameters $\lambda$ and $\delta$ require prior specification, where $\lambda \in (0,1)$ represents the internal-external data consistency (compatibility) threshold, governing the activation of external data borrowing. When $p \geq \lambda$, the system employs the vague prior with the minimally necessary weight $\omega_{Eq}$ (i.e., assigning maximal weight to external data). If $p$ remains below $\lambda$ (indicating significant conflict between external and current data), $\omega_{Eq}$ is set to 1, resulting in exclusive reliance on the vague prior without external data borrowing. Since smaller $p$ values reflect stronger inconsistency between internal and external data, we recommend setting $\lambda$ to a relatively large value to ensure borrowing is initiated only when external data meet predefined compatibility criteria.

The second parameter $\delta$, serves as a clinically defined equivalence margin, quantifying the permissible deviation between the mixed posterior (integrating external data) and the current trial posterior. A symmetric equivalence interval $[p_{Hyb} - \delta, p_{Hyb} + \delta]$ is constructed around the mixed posterior estimate $p_{Hyb}$, and the probability $Pr\ (p_{Hyb} - \delta < p_{Curr} < p_{Hyb} + \delta)$ is calculated to assess clinical equivalence. The value of $\delta$ must align with clinical consensus; for instance, efficacy endpoints may tolerate larger $\delta$ values compared to safety endpoints, thereby accommodating context-specific medical requirements for equivalence margins.

The parameters $\lambda$ and $\delta$ exhibit synergistic roles: $\lambda$ acts as a statistical threshold to restrict data borrowing under incompatibility, while $\delta$ introduces clinical flexibility to calibrate equivalence standards. This dual mechanism balances statistical rigor with enhanced adaptability to Real-world data complexity. Detailed operationalization of these parameters and their empirical impacts will be systematically examined in subsequent simulation trials.

*Step 4: Posterior distribution*

For binary endpoints, the posterior distribution is

$$g_{EQPS}(\theta_{curr}|d_{Real}, d_{Ex}, d_{Curr})$$

$$= (1 - \hat{\omega}_{Eq}) \sum_{k=1}^{K} \hat{\pi}_k Beta(a_k + y_{Curr}, b_k + N_{Curr} - y_{Curr})$$

$$+ \hat{\omega}_{Eq} Beta(a_0 + y_{Curr}, b_0 + N_{Curr} - y_{Curr}) \#(2-20)$$

Among them, in order to flexibly adapt to the problem of consistency change due to the increase or decrease of data in new areas, this paper adopts the update of the weights $\hat{\omega}_{Eq}$ and $\hat{\pi}_k$ here.

$$f_0 = \frac{B(a_0 + y_{Curr}, b_0 + N_{Curr} - y_{Curr})}{B(a_0, b_0)}$$

$$f_k = \frac{B(a_k + y_{Curr}, b_k + N_{Curr} - y_{Curr})}{B(a_k, b_k)}$$

$$\hat{\omega}_{Eq} = \frac{\omega_{Eq} f_0}{(1 - \omega_{Eq}) * \sum_k \omega_k f_k + \omega_{Eq} f_0}$$

$$\hat{\pi}_k = \frac{\pi_k f_k}{\sum_k \pi_k f_k} \#(2-21)$$

Overall, the figure illustrates the bridging strategy between the test drug group and the control drug group for the bridging trial in the new region. The outcome is presented as a binomial distribution for simplicity, but other common distributions such as continuous type distribution and TTE can be used as needed.

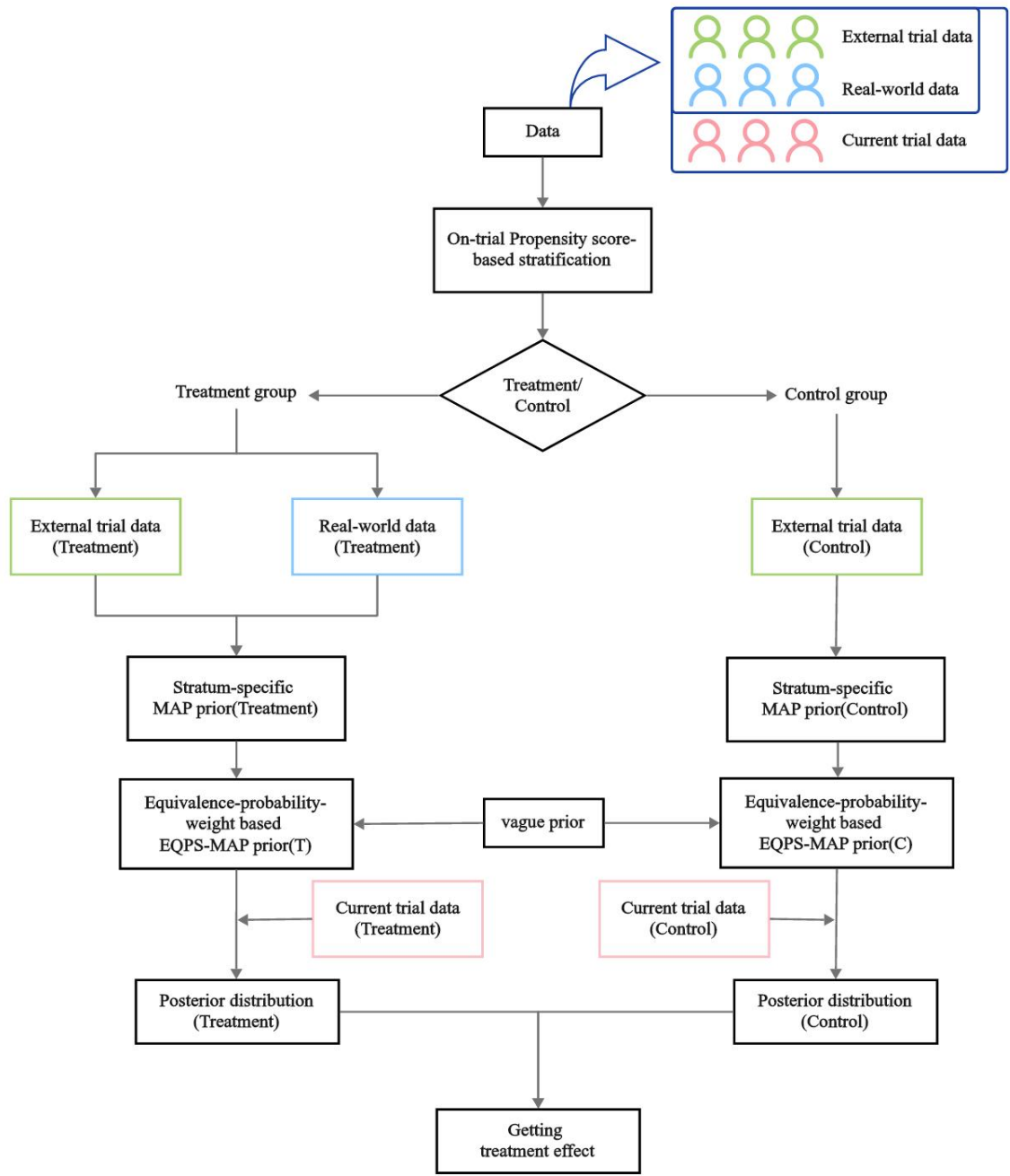

**Figure 2** Flowchart of the methodological organizing framework

## 3 Simulation study

To examine the performance of EQPS-rMAP, this section simulates various parameter analyses.

*3.1 Parameters specification*

Consider a typical two-arm randomized trial in a new region with 500 patients in each of the treatment and control groups, using a unilateral Type I error rate of 5% and a binary endpoint. There are External trial data with 500 patients enrolled in both the treatment and control group, and Real-world data with 500 patients enrolled in the treatment group. A multinomial logistic model was used to model the groups to which different baseline covariates belonged.

$$p(Z_1 = 1|X_i) = \frac{e^{\beta_{Real}*X}}{1 + e^{\beta_{Real}*X} + e^{\beta_{EX}*X}}$$
$$p(Z_2 = 1|X_i) = \frac{e^{\beta_{EX}*X}}{1 + e^{\beta_{Real}*X} + e^{\beta_{EX}*X}} \#(3-1)$$

$Z_{1i}$ is Real-world data indicator, and $Z_1 = 1$ indicates that subjects were from Real-world data. $Z_{2i}$ is External trial data indicator, and $Z_2 = 1$ indicates that subjects were from External trial data. $\beta_{Real}$ and $\beta_{EX}$ show the imbalance in baseline covariates between Real-world data and Current trial data, and between External trial data and Current trial data, respectively. Individual outcomes were generated by simulation in the model.

$$logit(Y_i = 1|X_i, T_i, Z_i) = \beta_0 + \beta_1 T_i + \beta_2^T X_i + \beta_3 T_i Z_{1i} + \beta_4 T_i Z_{2i} \#(3-2)$$

$T_i$ is the treatment indicator. $\beta_1$ represents the treatment effect. $\beta_2^T$ represents the impact of covariates on the outcome, while $\beta_3$ represents effect heterogeneity between Real-world data and current trial data. $\beta_4$ represents effect heterogeneity between External trial data and current trial data.

In the simulation experiments, $X$ is set to a combination of binary and continuous variables for simplicity and $\beta_0 = 0, \beta_1 = 0.5, \beta_2^T = 0.5$ . The hypotheses are $H_0: \theta_c = \theta_t$ against $H_1: \theta_c < \theta_t$, where $\theta_c$ and $\theta_t$ are the objective response rates for the control and treatment groups, respectively. The trial is considered statistically successful if the posterior probability of treatment superiority exceeds a pre-specified 95% Bayesian threshold (equivalent to controlling the Type I error rate at 5% under frequentist calibration):

$$P(\theta_t > \theta_c | External\ trial, Real-world\ and\ Current\ trial\ data) > 0.95.$$

The simulation study focuses on two objectives: (1) validation of the feasibility of EQPS-rMAP, and (2) evaluation of statistical performance of EQPS-rMAP. To validate feasibility, we examined the changes in mixed prior weights ($\widehat{\omega}_{Eq}$) and the sample size ratio ($N_{\text{EQPS−rMAP borrowing}}/N_{\text{no−borrowing design}}$) to assess the adaptive capability of EQPS-rMAP across various scenarios. These scenarios consisted of two levels of baseline differences, three levels of effect heterogeneity, and nine combinations of the parameters λ and δ, resulting in a total of 54 scenarios (2 baseline differences×3 effect heterogeneity×3λ× 3δ). For the evaluation of statistical performance, EQPS-MAP ($\lambda = 0.8, \delta = 0.1$) was compared with traditional methods (EB-rMAP, MAP, and PS-MAP) using absolute bias, mean squared error (MSE), and type I error with regarding to $\beta_1$. The comparison was conducted under six scenarios defined by two levels of baseline differences and three levels of effect heterogeneity.

The simulation experiments used the vague prior $Beta(1,1)$. Markov Chain Monte Carlo (MCMC) calculations were implemented using R 4.2.3 with Rstan. For all methods, we started five chains and ran 41,000 MCMC iterations with a burn-in period of 1000.

*3.2 Simulation results*

The simulation results are presented in a series of plots that illustrate the relationship between the heterogeneity of the RWD with the current study data and the weight of the hybrid prior, where a larger weight indicates increased use of External data.

Figures 3 and 4, each with lines for different values of parameter $\lambda$, visualize the effect of the heterogeneity variance $\delta$(0, 0.2, 0.4) and the values of parameter $\delta$ (0.1, 0.15, 0.2) on the weights of the hybrid prior. The impact of varying parameter values on the weights of the mixed prior under varying levels of heterogeneity is fully elucidated. The curves exhibit a general tendency of increasing and then decreasing. This is due to the fact that the weight of the mixed prior increases when the horizontal coordinate is going to 0, indicating that the borrowed data are more consistent with the data of the new area. In all plots, the larger the weight threshold parameter $\lambda$, the more concentrated the larger curve. This is due to the fact that the larger the threshold, the more cautious the borrowed data is, and the range of intervals that can be borrowed becomes concentrated. A comparison of the three columns from left to right demonstrates that the larger the tolerable error parameter $\delta$, the wider either curve is, indicating that the more relaxed the requirements are, and the more information can be borrowed with equal heterogeneity. The conflict minimization point is defined as the heterogeneity parameter combination ($\beta_3, \beta_4$) that minimizes discordance between External borrowable data (External trail data and Real-world data) and Current trial data, corresponding to the peak of mixed prior weights (maximum borrowing efficiency) and the trough of required Current trial sample size as a proportion of the original (minimum sample burden) in the figure. The curve's central axis rightward shift reflects heterogeneity-driven displacement of this minimization point. Specifically, opposing-direction heterogeneities with equal absolute magnitudes($\beta_3 = -\beta_4$) induce partial cancellation through weighted prior effects, enhancing alignment with current trial data and thereby minimizing composite prior-data conflict.

At the same time, the curve does not become more centralized as a result of the heterogeneity difference. This is because our method is capable of handling the heterogeneity simultaneously without affecting the scope of the borrowed information. For the cases with and without baseline differences, the conclusions remain essentially the same. For the case of the presence of baseline differences, the curves are slightly more concentrated. The presence of baseline differences affects the size of data similarity to some extent, which in turn results in a reduction in the amount of borrowed information that can be applied under the same rule.

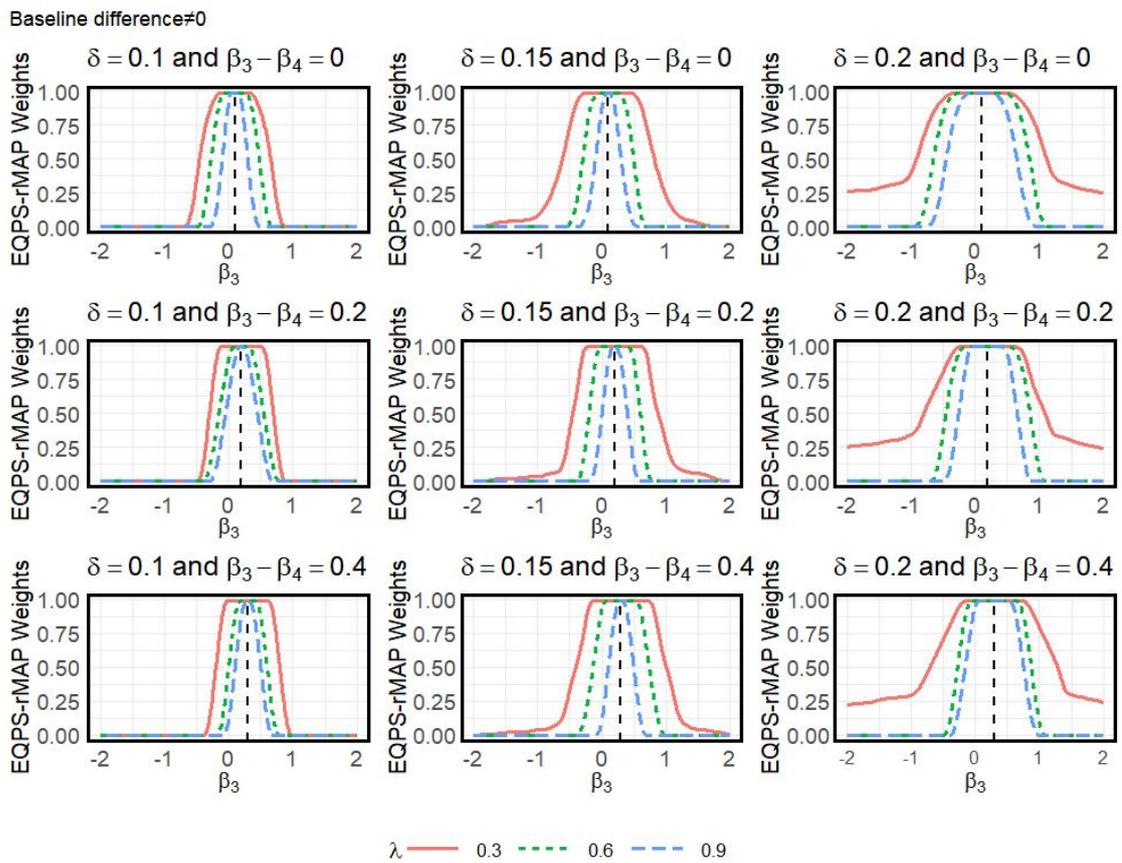

**Figure 3** Combinatorial plots of the change in prior weights with heterogeneity for different

parameter values and different heterogeneity differences in the absence of baseline differences.

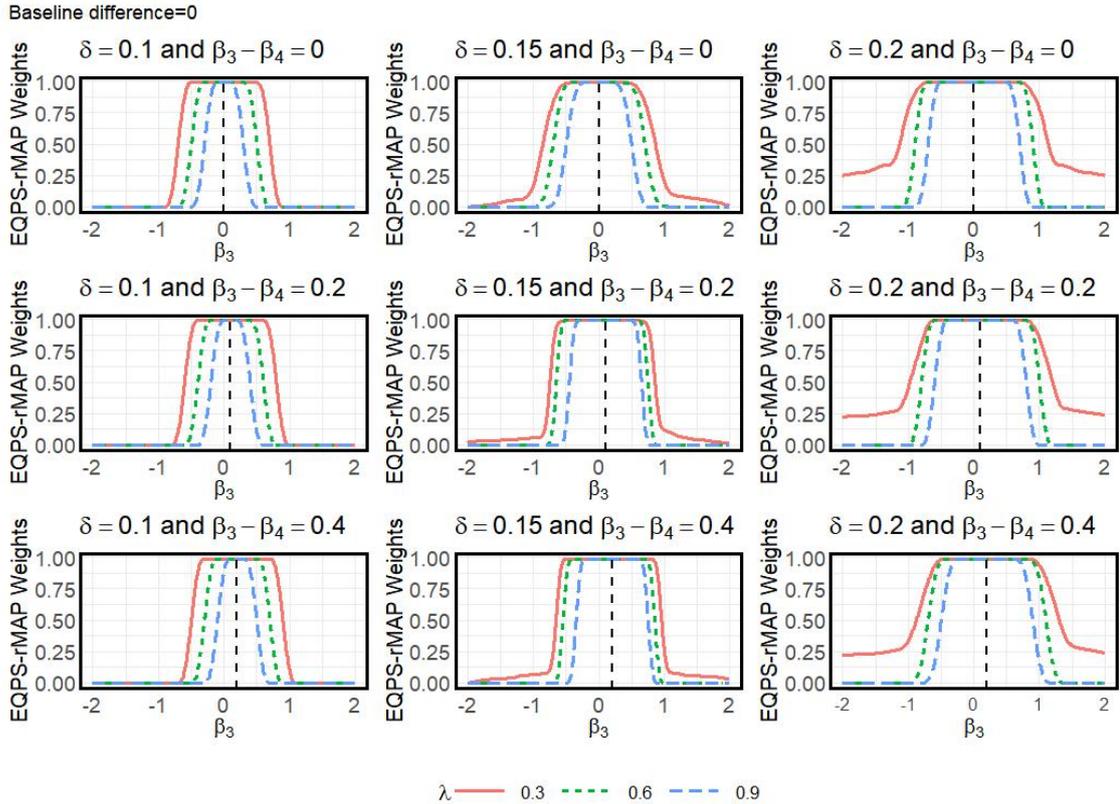

**Figure 4** Combinatorial plots of the change in prior weights with heterogeneity for different parameter values and different heterogeneity differences in the presence of baseline differences.

The method proposed in this paper also makes it very simple to calculate how much sample size was borrowed. Figure 5 and 6 illustrate the relative sample size required for statistical inference in new regions compared to the original size required. These figures effectively show how different parameter values affect the required sample size under varying degrees of heterogeneous variance. In contrast to the observed weight trends, the curves exhibit a downward and then upward trajectory. Consequently, the conclusions regarding the threshold parameter $\lambda$, the allowable error parameter $\delta$, and the curve shift remain largely consistent with the weighting conclusions. Furthermore, a comparison of the three rows from the top to the bottom reveals that the greater the level of heterogeneous variation, the greater the minimum value of the curve required for the new area, indicating that less information is borrowed. The presence of baseline differences also results in a reduction in the amount of borrowed information compared to the absence of baseline differences under the same rule. This is due to the fact that both the level of heterogeneity and the baseline differences influence the degree of data similarity to some extent.

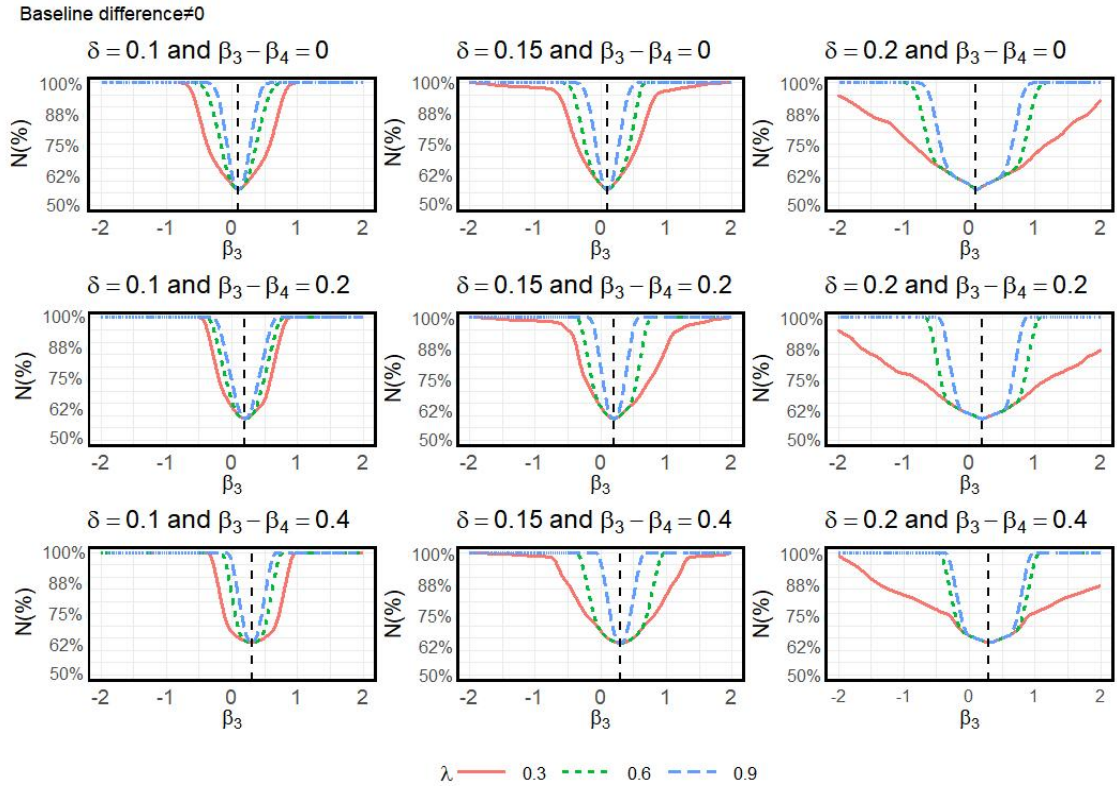

**Figure 5** Combined plots of the change in sample size required to reach statistical inference with heterogeneity for new areas for different parameter values and different heterogeneity differences in the absence of baseline differences.

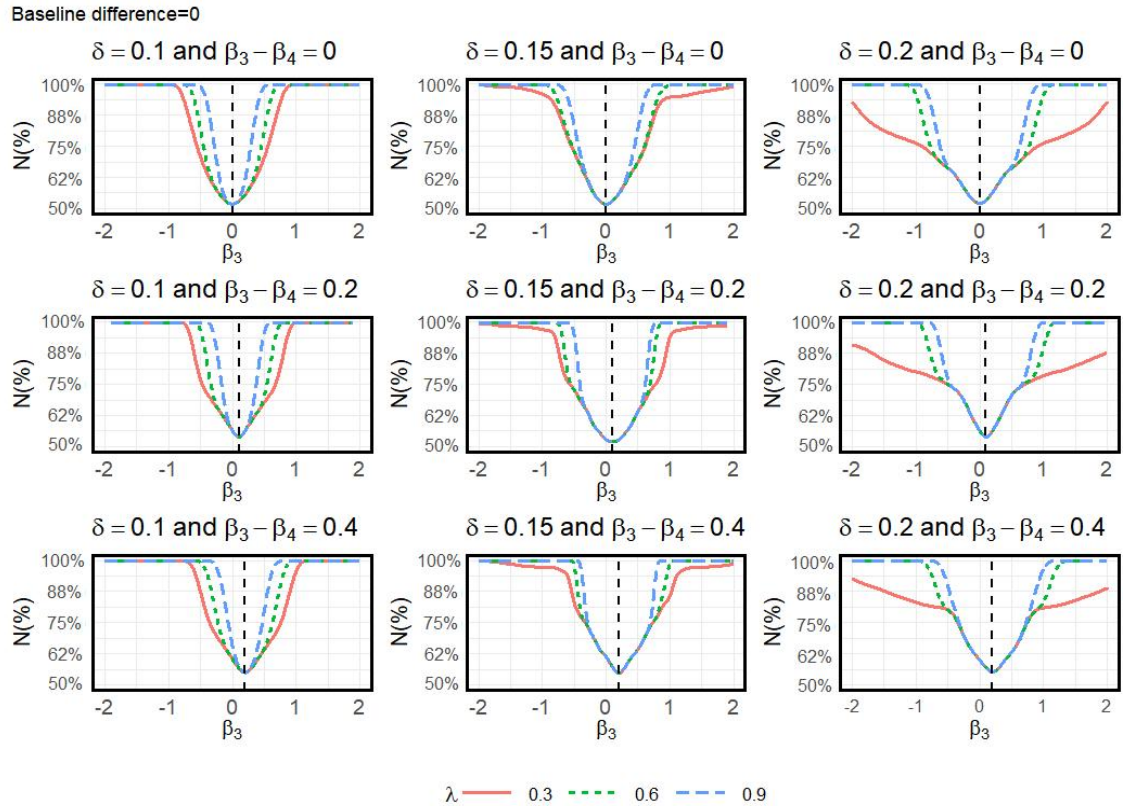

**Figure 6** Combined plots of the change in sample size required to reach statistical inference with heterogeneity for new areas for different parameter values and different heterogeneity differences in the presence of baseline differences.

Figures 7 and 8 compare the EQPS-rMAP method with the EB-rMAP and PS-MAP methods using bias, mean squared error (MSE), and Type I error as metrics. EQPS-rMAP demonstrates superior performance, maintaining low bias and MSE while effectively managing Type I error, regardless of baseline differences. When there is no baseline difference and heterogeneous difference, the proposed method EQPS-rMAP is close to the EB-rMAP method and PS-MAP is close to the MAP method. The larger the heterogeneity difference is, the advantage of the proposed method EQPS-rMAP is more obvious, because this method does not borrow data in a blended way like previous methods. EQPS-rMAP is superior to EB-rMAP when the baseline difference is not 0 compared to the baseline difference is 0. The advantage of PS-MAP is greater than MAP.

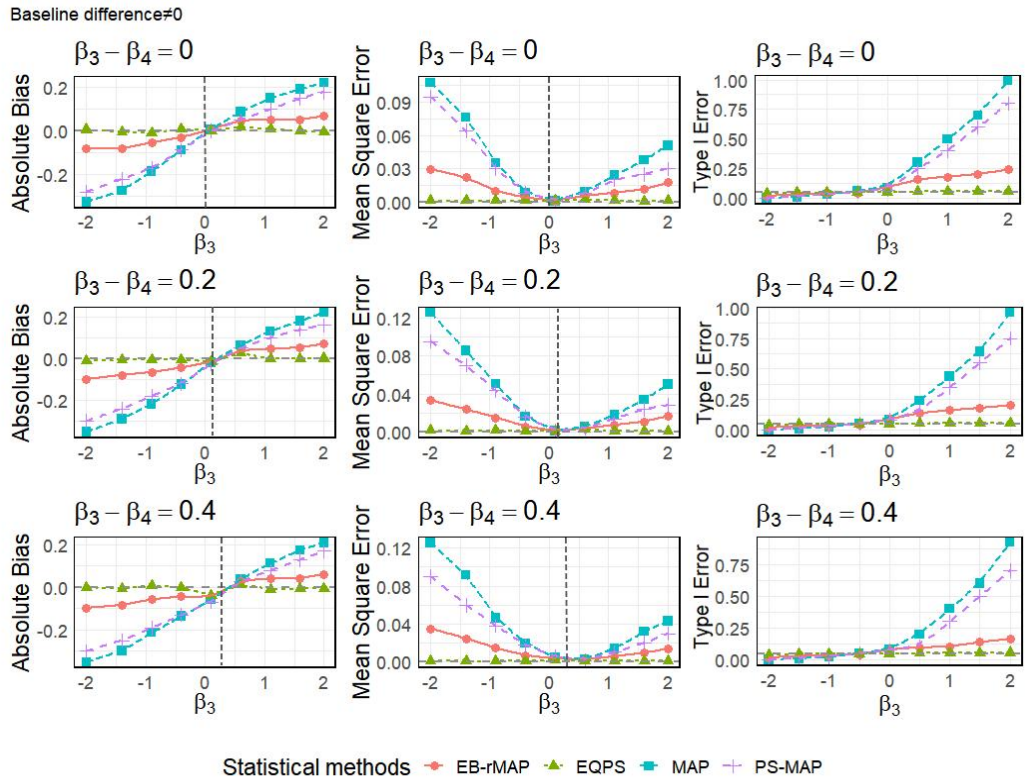

**Figure 7** Mean Square Error, Absolute bias and Type I Error for EQPS-rMAP, MAP, PS-MAP, EB-rMAP in the absence of baseline differences.

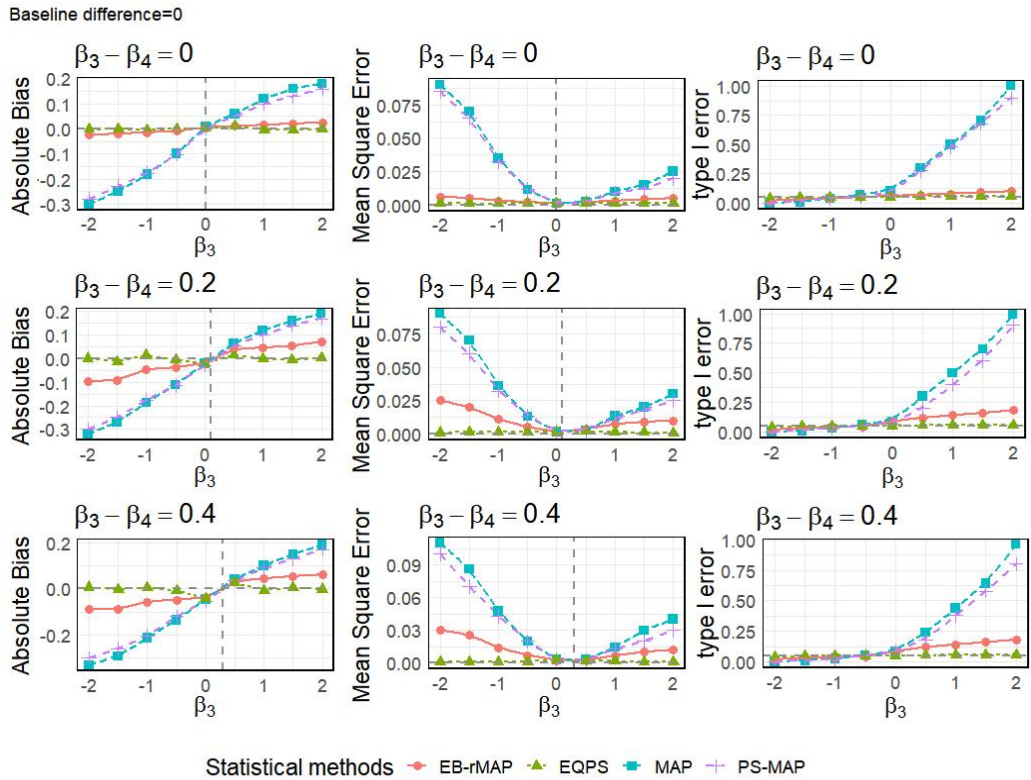

**Figure 8** Mean Square Error, Absolute bias and Type I Error for EQPS-rMAP, MAP, PS-MAP, EB-rMAP in the presence of baseline differences.

## 4 An illustrative example

This section validates the applicability of the proposed methodology through a clinical case study. We evaluated the efficacy of risankizumab in patients with moderate-to-severe plaque psoriasis using two independent datasets: the UltIMMa-1/UltIMMa-2 pooled analysis (Gooderham M et al., 2022)[36] as a randomized controlled trial (RCT) data source and a 40-week multicenter retrospective cohort reported by Borroni RG et al. (2021) [37]. Since risankizumab is not yet approved by Chinese regulatory authorities and real-world data (RWD) from Hainan Lecheng are still under collection, External trial data (UltIMMa-1/UltIMMa-2) and Real-world data (a retrospective treatment cohort, 40-week follow-up) were utilized to construct external borrowable data, with Current trial data generated through simulation. Specifically, hypothetical Current trial data datasets were simulated based on baseline characteristics from External trial data, with predefined response rates for treatment and control groups informed by External trial data and Real-world data discrepancies.

Due to restricted access to patient-level data from UltIMMa-1/UltIMMa-2 and the retrospective cohort (40-week follow-up), we simulated these datasets using aggregate baseline statistics reported in published literature to ensure reproducibility. The primary endpoint was defined as the proportion of patients achieving an absolute PASI threshold (PASI$\leq$1) at week 40 of treatment, with hypothesis testing framed as superiority testing ($H_0: \theta_c = \theta_t$ against $H_0: \theta_c < \theta_t$), where $\theta_c$ and $\theta_t$ denote objective response rates (ORR) for control and treatment groups, respectively. In the original data, the UltIMMa-1/UltIMMa-2 trials included 598 patients (399 in the treatment group, 71.9%; 199 in the control group, 38.7%), while the retrospective cohort comprised 77 treated patients (85.7%). For a hypothetical scenario requiring additional RCTs for risankizumab's approval in China, we constructed a simulated dataset (100 patients per group) with a 40% response rates in control group and 65% in the treatment group.

We used IPW to control for differences in the following baseline factors: age, sex, BMI, PASI, Exposure to previous biologics, Anti-TNF, Anti-IL 17. The analysis of the example data was realized in R 4.2.3.

Figure 9 illustrates the probability density distributions of treatment groups across data sources at the 40-week treatment endpoint, including the current trial data (no data borrowing), external trial data prior (UltIMMa-1/UltIMMa-2), real-world data prior, and the posterior distribution generated by the proposed EQPS-rMAP method. Significant heterogeneity discrepancies are observed between borrowable external data (UltIMMa-1/UltIMMa-2 trials and Real-world data) and current trial data, where conventional static borrowing approaches would introduce bias in posterior estimation. The dynamic Bayesian borrowing method addresses this limitation by quantifying data conflicts and selectively integrating high-consistency external data sources. This strategy enhances research efficiency through external information incorporation (evidenced by significantly increased peak amplitude, confirming external data utilization) while maintaining robust estimation accuracy—the posterior distribution converges centrally near the true efficacy value. These results demonstrate that EQPS-rMAP's heterogeneity-adaptive mechanism effectively identifies and retains external evidence compatible with current trial data, achieving an optimal balance between bias control and information gain.

Figure 10 presents a comparative analysis of posterior estimation trends between traditional methods (MAP, EB-rMAP, and PS-MAP) and the proposed EQPS method across varying proportions of borrowed data. The dashed vertical line at $X = 0.25$ marks the prespecified true treatment effect for the current region , serving as a reference for evaluating estimation accuracy. Notably, the results demonstrate that EQPS maintains stable and accurate estimations regardless of data borrowing ratios, whereas traditional methods exhibit significant variability in performance. The proportion of borrowed data may be varied by changing the sample size of the RWD data in the table. As the proportion of Externally borrowed data increases, traditional methods are affected and estimates are shifted. This is a common problem with traditional methods such as MAP and PS-MAP, where data from a new area cannot reverse the results of External data. Additionally, even with its improved robustness, EB-rMAP is affected by baseline differences in data components. Given that the impact of prior-data conflicts is less pronounced than the baseline discrepancies observed in the example, the efficacy of the methodologies can be ranked in the following order: EQPS > PS-MAP > EB-rMAP > MAP.

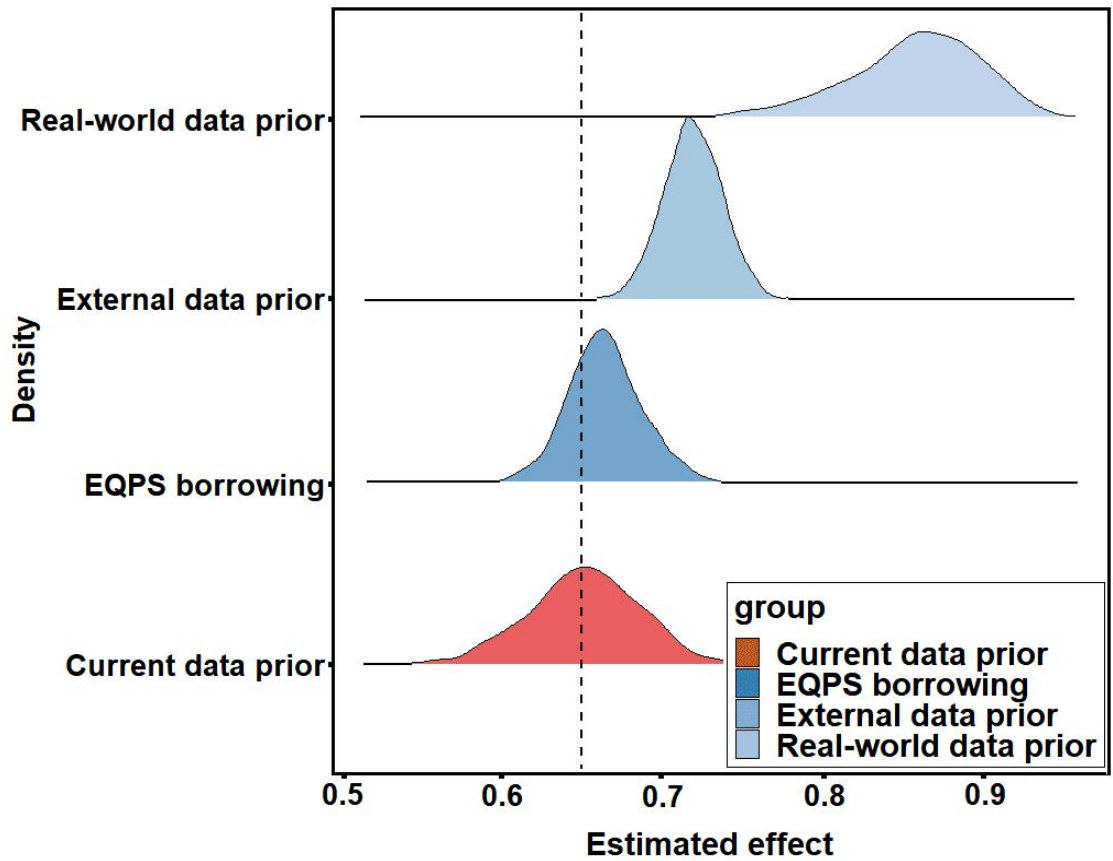

**Figure 9** Probability density distribution of $\theta_t$, compared with those of current trial data and EQPS, External data prior, Real-world data prior with normally distributed outcomes.

Note：

[1] Current trial data：represents the distribution of the current trial data when no external data is borrowed.

[2] Dashed line: represents the true efficacy of the current trial data, $x = 0.65 - 0.4$;

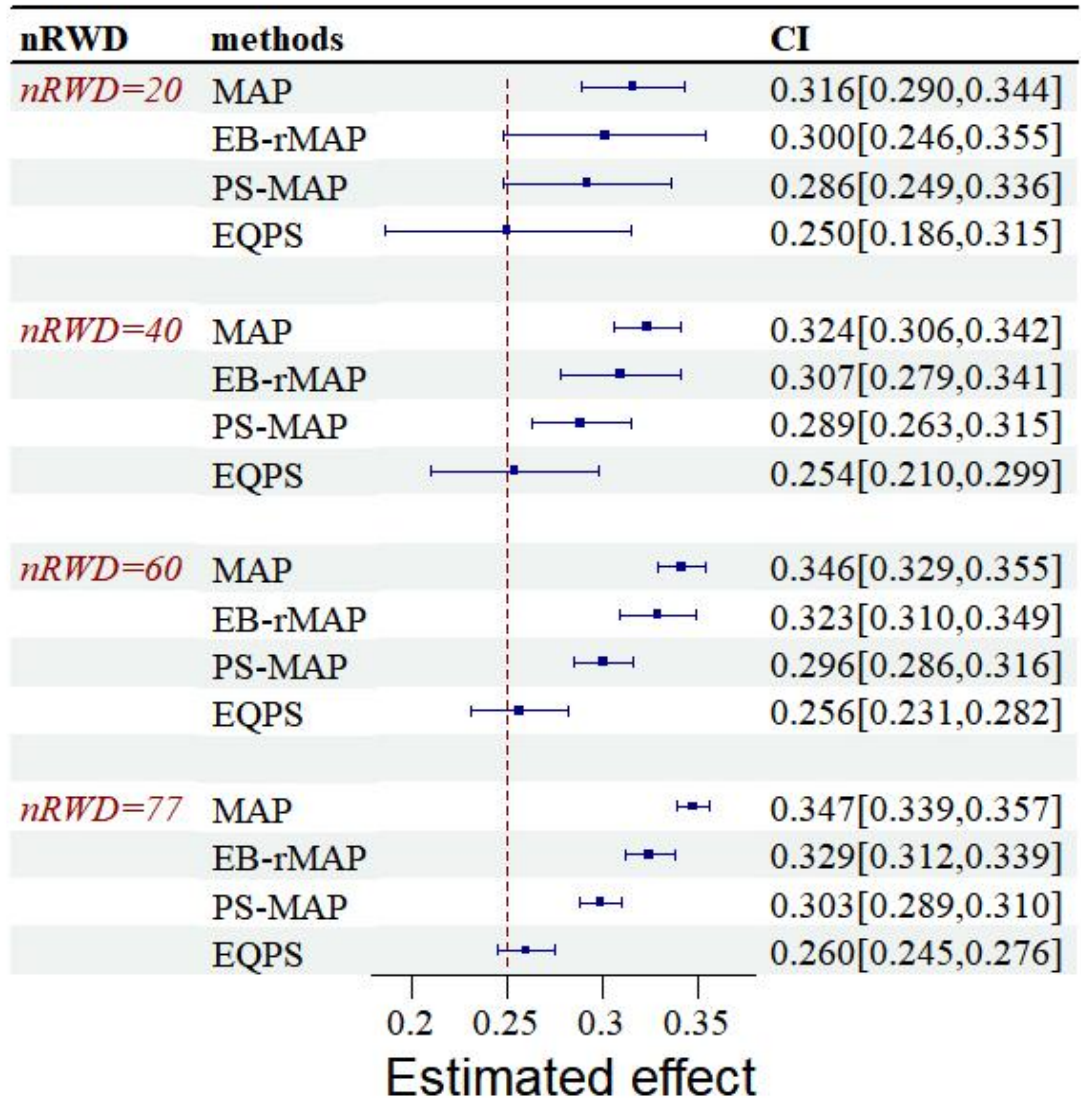

**Figure 10** The posterior distribution of $(\theta_t - \theta_c)$, compared with MAP, PS-MAP, EB-rMAP

with normally distributed outcomes. The dashed line at X=0.25 indicates the prespecified true treatment effect for the current region.

## 5 Conclusion

In March 2024, the U.S. Food and Drug Administration (FDA) convened a workshop titled "Advancing Complex Innovative Trial Designs in Clinical Trials: From Pilot to Practice" and announced plans to release draft guidance on the application of Bayesian methods in clinical trials by the end of 2025 (FDA, 2023i)[38]. The workshop emphasized the integration of external data sources, Bayesian statistical techniques, and simulation technologies in complex innovative trial designs (FDA, 2024a)[39]. This paper proposes the EQPS-rMAP method, which employs a hybrid design to eliminate baseline differences between groups and inter-study heterogeneity, aiming to optimize the integration of External trial data and Real-world data. This approach provides an innovative solution to accelerate drug development and significantly enhance the efficiency of randomized controlled trials (RCTs).

Systematic simulation experiments validated the robustness of the dynamic Bayesian borrowing method across diverse scenarios, including extreme conditions.

Compared to conventional methods such as MAP and PSMAP—which allow predefined borrowing proportions but fail to control bias under increasing sample sizes due to fixed full-sample borrowing strategies—the proposed method addresses critical limitations. Traditional EB-rMAP approaches, while dynamically adjusting borrowing scales based on data similarity[22], assume homogeneity across data sources. This assumption risks bias under significant baseline or effect heterogeneity, leading to overly conservative borrowing (complete exclusion of external data) or erroneous borrowing (introducing highly biased data). The EQPS-rMAP method overcomes these challenges through a two-stage optimization: First, propensity score stratification resolves baseline discrepancies between RWD and trial data40, ensuring balance in key predefined covariates. Second, a differentiated weighting strategy addresses multi-source heterogeneity, establishing an adaptive borrowing mechanism based on quantified prior-data conflicts. Compared to EB-rMAP[22], EQPS-rMAP introduces additional tuning parameters to enhance flexibility, though systematic simulations are required to identify optimal parameter combinations. This framework is applicable not only to bridging studies but also to extension strategies in multi-regional clinical trials (MRCTs) and post-marketing confirmatory research.

However, three limitations remain: First, the current framework supports only qualitative efficacy analysis (determining whether the difference between new-region treatment and control groups exceeds 0) without quantifying the magnitude of differences[12]. A practical solution involves comparing efficacy differences to predefined non-inferiority margins[41], which can be seamlessly integrated into the existing framework. Second, while the study focuses on improving research efficiency and estimation accuracy, it has yet to explore patient benefit optimization through adaptive designs (e.g., dynamic randomization). Future work could integrate patient preference models to develop a multidimensional benefit evaluation system[42-43]. Third, while the method performs robustly for binary endpoints, it requires extension to complex endpoints such as survival analysis. Developing time-to-event frameworks is critical for broader clinical applicability. Furthermore, MRCT extension strategies lack operational guidelines; collaboration with regulators is needed to establish scientific criteria for prior data borrowing and efficacy thresholds, ensuring methodological implementation.

**Data accessibility statement**

The data and R code to replicate the data analysis in Section 4 are available.

**Funding**

Ying Wu is supported by the National Natural Science Foundation of China [Grant number 82273732], the Guangzhou Basic and Applied Foundation Project (2023A04J1106) and the Real World Research Project Grant Fund from the Hainan Institute of Real World data (HNLC2022RWS018).

**Acknowledgements**

The authors thank William Wang, Helen Wu, and Weiwei Zhao from the SMU-MSD BARDS Academic Projects of Merck for their valuable comments.

**Declaration of conflict of interests**

The Authors declare that there is no conflict of interest.